\documentclass[%
reprint,
 amsmath,amssymb,
 aps,
]{revtex4-1}

\usepackage{graphicx}
\usepackage{dcolumn}
\usepackage{bm}

\begin{document}

\preprint{APS/123-QED}

\title{Nuclear matrix elements for  neutrinoless $\beta \beta $ decays and  spin-dipole giant resonances}

\author{Hiroyasu Ejiri}
\email{ejiri@rcnp.osaka-u.ac.jp}
 \affiliation{Research Center for Nuclear Physics, Osaka University, Osaka 567-0047, Japan
 }%
 
\author{Lotta Jokiniemi}%
\affiliation{Department of Quantum Physics and Astrophysics and Institute of Cosmos Sciences,  University of Barcelona, Barcelona 08028, Spain
}%
\author{Jouni Suhonen}
\affiliation{Department of Physics, University of Jyvaskyla, Jyvaskyla,
FI-40014, Finland
}%
\date{\today}
\begin{abstract}
Nuclear matrix element (NME) for neutrinoless double beta decay (DBD) is required for studying  neutrino physics beyond the standard model by using DBD. 
 Experimental information on nuclear excitation and decay associated with DBD is crucial for theoretical calculations of the DBD-NME. 
The spin-dipole ({\rm SD}) NME for DBD via  the intermediate SD state  is  one of the major components of the DBD-NME. 
 The experimental  SD giant-resonance energy and the SD strength  in the intermediate nucleus  are shown for the first time to be closely related to the DBD-NME and  are used for studying the  spin-isospin correlation and the  quenching of the axial-vector coupling, which are involved in the NME. So they are used to help the theoretical model calculation of the DBD-NME. 
  Impact of the  SD giant-resonance and the SD strength on the DBD study is discussed.  
\end{abstract}

\maketitle



Neutrinoless double beta decay (DBD),
which violates the lepton-number conservation law,
 is a sensitive and realistic probe for studying the neutrino ($\nu$) nature (Majorana or Dirac) relevant to the origin of matter in the universe, the absolute $\nu$-mass scale, and other $\nu$-properties beyond the standard model \cite{eji05,avi08,ver12}.  The nuclear matrix element (NME)  for the neutrinoless DBD is crucial to extract the effective $\nu$-mass and other $\nu$-properties of the particle physics interests from the decay rate.  The NME is also needed to design DBD detectors  \cite{eji78,eji00,eji19}. Thus the NMEs are of great interest from astro-, particle- and nuclear-physics view points.  Theoretical works on DBD-NMEs by using various nuclear models are discussed in \cite{suh98,fae98,suh12,bar13,eng17,jok18}, and recent DBD experiments
 in \cite{eji20,det20,eji21a}.
 
 Actually, accurate theoretical calculations for the DBD-NME  are very hard since they are  sensitive to nucleonic and non-nucleonic  correlations and nuclear medium effects. Consequently, calculated DBD-NMEs, including the effective axial-vector coupling ($g_{\rm A}$), scatter over an order of magnitude \cite{ver12,eji19}, depending on the nuclear models, the interaction parameters and the effective coupling ($g_{\rm A}^{\rm eff}$) used in the models.
  Thus experimental inputs are useful to check the theoretical models and  the nuclear parameters to be used for the models \cite{eji05,eji19}.  
  
 The present letter aims to show for the first time that the experimental energy and the strength of the spin-dipole (SD) giant-resonance in the intermediate nucleus  are closely related to the DBD-NME ($M^{0\nu}$) based on the pnQRPA (proton-neutron Quasi-particle Random Phase Approximation) model, and 
	 reflect the nuclear structure and the quenching  of the $g_{\rm A}$, which are involved in the DBD-NME.  Here the SD strength is given as $B^-$(SD)=$|M^-$(SD)$|^2$ with $M^-$(SD) being the SD NME. We note that $M^-$(SD) is  associated with the SD component of the DBD-NME, which is one of the major  components of the DBD-NME, and the quenching of $g_{\rm A}$ is one of key parameters for the theoretical model calculation. Thus experimental information on the SD giant-resonance energy $E_{\rm G}$(SD) and the SD strength $B^-$(SD) are used to help and check the theoretical model calculation of the DBD-NME.
 
 Recently, the experimental $E_{\rm G}$(SD) values were shown to depend on the isospin $z$-component  $T_z$=($N-Z$)/2 with $N$ and $Z$ being the neutron and proton numbers \cite{eji19}, and  pnQRPA calculations for $M^{0\nu}$ were performed by adjusting the particle-hole parameter to the $E_{\rm G}$(SD) in \cite{jok18}, where $g_{\rm A}^{\rm eff}$=1 is assumed.   
Single-$\beta$ SD NMEs for the ground states in the medium-heavy nuclei (mostly non-DBD nuclei) were studied in the framework of the pnQRPA, and were found to be reduced by $g_{\rm A}^{\rm eff}$/$g_{\rm A}\approx$ 0.5, depending much on the individual states \cite{eji14}.
 
 The present paper  puts emphasis on studying and discussing universal features in the $E_{\rm G}$(SD), the GT and SD strengths, and the $M^{0\nu}$ as a function of the mass number $A$. For $M^{0\nu}$, we use a common value for the $g_{\rm A}^{\rm eff}$ as derived by referring to the experimental summed GT strengths for the DBD nuclei and the SD NMEs for the ground state transitions in the medium-heavy nuclei. Then we discuss impact of the present findings on the DBD experiments.

\begin{figure}[ht]
\hspace{0cm}
\vspace{-0cm}
\includegraphics[width=0.5\textwidth]{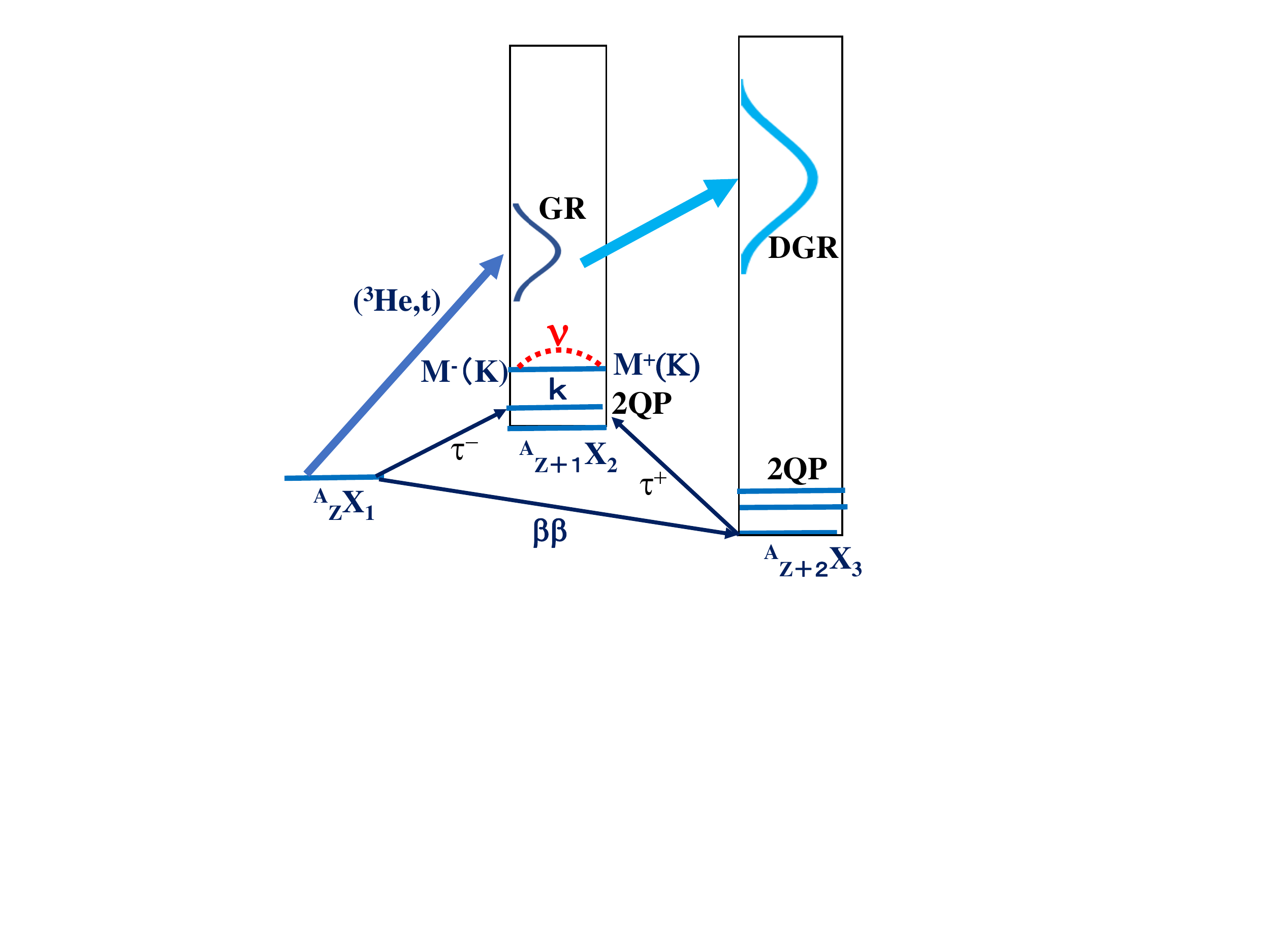}
\vspace{-3cm}
\caption{DBD transition scheme for $^A_{Z}$X$_1\rightarrow ^A_{Z+2}$X$_3$ 
 with the Majorana $\nu$ exchange in the intermediate nucleus  $^A_{Z+1}$X$_2$. $\tau^+$ and $\tau^-$: Isospin raising and lowering operators.
$M^-$(K) and $M^+$(K): $\tau^-$ and $\tau^+$  K-mode NMEs associated with the K-mode  DBD NME. k: Intermediate state. ($^3$He,$t$): Charge-exchange reaction.
2QP: Two-quasiparticle state.
 GR: Giant resonance.  DGR: Double giant resonance. See text.
\label{figure:fig1}} 
\end{figure}

We discuss  double $\beta ^-$ decays of the ground-state-to-ground-state transition of 0$^+\rightarrow 0^+$. The neutrinoless DBD (0$\nu\beta \beta $)   for  $^A_Z$X$_{1}
\rightarrow ^A_{Z+2}$X$_3$  is schematically shown in Fig.\ref {figure:fig1}. The intermediate nucleus is $^A_{\rm Z+1}$X$_2$. The 0$\nu\beta \beta $ process to be discussed is the  Majorana $\nu$-mass process of the current interest. 
Here,  a light Majorana $\nu$ is virtually exchanged  between two neutrons in the DBD nucleus. DBD nuclei discussed  are medium-heavy ($A$=76-136) nuclei with the high $\nu$-mass sensitivity \cite{eji20}.

The DBD rate is expressed as \cite{eji05,ver12,eji19}
\begin{equation}
R^{0\nu}={\rm ln2}~g_{\rm A}^4G^{0\nu}[m_{\beta\beta} |M^{0\nu}|]^2,
\end{equation}
where $G^{0\nu}$ is the phase space factor, $g_{\rm A}$=1.27 is the axial-vector coupling for a free nucleon in units of the vector coupling of $g_V$, $m_{\beta\beta}$ is the effective $\nu$-mass and  $M^{0\nu}$ is the DBD NME.   
The NME is given as \cite{ver12,eji19,suh98,jok18}
\begin{equation}
M^{0\nu}=(\frac{g_{\rm A}^{\rm eff}}{g_A})^2 [M^{0\nu}_{\rm GT} + M^{0\nu}_{\rm T}] -  (\frac{g_V}{g_A})^2M^{0\nu}_{\rm  F},
\end{equation}
where $M^{0\nu}_{\rm GT}$, $M^{0\nu}_{\rm T}$  and $M^{0\nu}_{\rm F}$ are the axial-vector (GT: Gamow-Teller), tensor (T) and vector (F: Fermi) DBD-NMEs, respectively, and $g_{\rm A}^{\rm eff}$ is the effective axial-vector coupling introduced to incorporate the quenching effect. 

DBDs involve mainly  axial-vector  spin-isospin ($\sigma \tau$) and vector isospin ($\tau$) transitions. The NMEs depend much on nucleonic and non-nucleonic $\sigma \tau$ and 
 $\tau$ interactions and their correlations. The pnQRPA model, which includes explicitly 
the nucleonic interactions and their correlations, has been widely used. Then the NMEs $M^{0\nu}_{\alpha}$ with $\alpha=$GT, T, F are given by the pnQRPA model NMEs, and $g_{\rm A}^{\rm eff}/g_{\rm A}$ stands for the re-normalization (quenching) coefficient due to the non-nucleonic $\sigma \tau$ correlations, nuclear medium effects and others which are not explicitly included in the pnQRPA model. Then  $g_{\rm A}^{\rm eff}$/$g_{\rm A}$  depends on the model to be used and the experimental data to be compared with, as discussed in \cite{eji78,eji19,eji14,suh17,suh17a,eji19a,sim13}. In particular, the present
 pnQRPA is the one which is standardly used in the beta and 
double-beta calculations, and it is explained in great detail 
in \cite{suh07}. The $g_{\rm V}$=1  for a free nucleon is assumed in Eq. (2).

The NME $M^{0\nu}_{\alpha}$ is given by the sum of the NMEs $M^{0\nu}_{\alpha}(k)$ for the intermediate states $k$ as \cite{ver12,eji19,suh98,jok18}
\begin{equation}
M^{0\nu}_{\alpha}=\sum_k M^{0\nu}_{\alpha}(k),~~~M^{0\nu}_{\alpha}(k)= <T_{\alpha}(k)>,
\end{equation}
where the transition operator $T_{\alpha}(k)$ with $\alpha$=GT,T and F are given by
$T_{\rm GT}(k)=
 t^{\pm}\sigma h_{\rm GT}(r_{12},E_k) t^{\pm}\sigma$,
$T_{\rm T}(k)=t^{\pm}h_{\rm T}(r_{12},E_k)S_{\rm 12}t^{\pm}$, and 
$T_{\rm F}(k)=t^{\pm}h_{\rm F}(r_{12},E_k)t^{\pm}$.
The operator includes the neutrino potential  $h_{\alpha}(r_{\rm 12},E_k)$  with $r_{\rm 12}$  being the distance between the two neutrons involved in the $\nu$ exchange and $E_k$ being the  excitation energy of the intermediate state $k$,  $S_{\rm 12}$ is the spin tensor operator and $t^{\pm}$ is the isospin operators for proton $\rightleftharpoons$neutron.
Since the  momentum of the virtual-$\nu$  is of the order of 1/$r_{\rm 12}\approx$100 MeV/c, the NME involves  mainly intermediate states $k$ in the wide ranges of the spins of $J^{\pi}\approx$$0^{\pm}-7^{\pm}$ and  $E_k\approx$ 0-30 MeV.  Thus, the DBD-NME may reflect gross properties of the nuclear core. 
 
The axial-vector SD ($J^{\pi}=2^-$) component of the $M^{0\nu}_{GT}$({\rm SD}) in Eq. (2) is one of  the major components since the orbital angular momentum matches the medium momentum of the virtual neutrino. The SD DBD-NME is associated with the product of the single $\tau^{\pm}$ SD NMEs of $M^+$({\rm SD}) and  $M^-$({\rm SD}) , as shown in Fig. 1. The single-$\beta$ SD NMEs are expressed as $M^{\pm}({\rm SD})$= $<t^{\pm} [\sigma f(r) Y_1]_2>$ with $Y_1$ being the spherical harmonics.

On the other hand, the two-neutrino DBD within the standard model is followed by the emission of the  two real  s-wave ($l$=0) neutrinos with low momentum, and the NME ($M^{2\nu}$) involves exclusively the  $M^{\pm}({\rm GT})$ for GT 1$^+$ intermediate states at the low excitation-energy region. Thus $M^{2\nu}$ is very sensitive to properties of the valence-nucleons at the proton neutron Fermi surfaces. 
So far,  most theoretical DBD models use these valence-nucleon GT properties.\\
\begin{figure}[htb]
\hspace{-0cm}
\vspace{-0cm}
\includegraphics[width=0.95\textwidth]{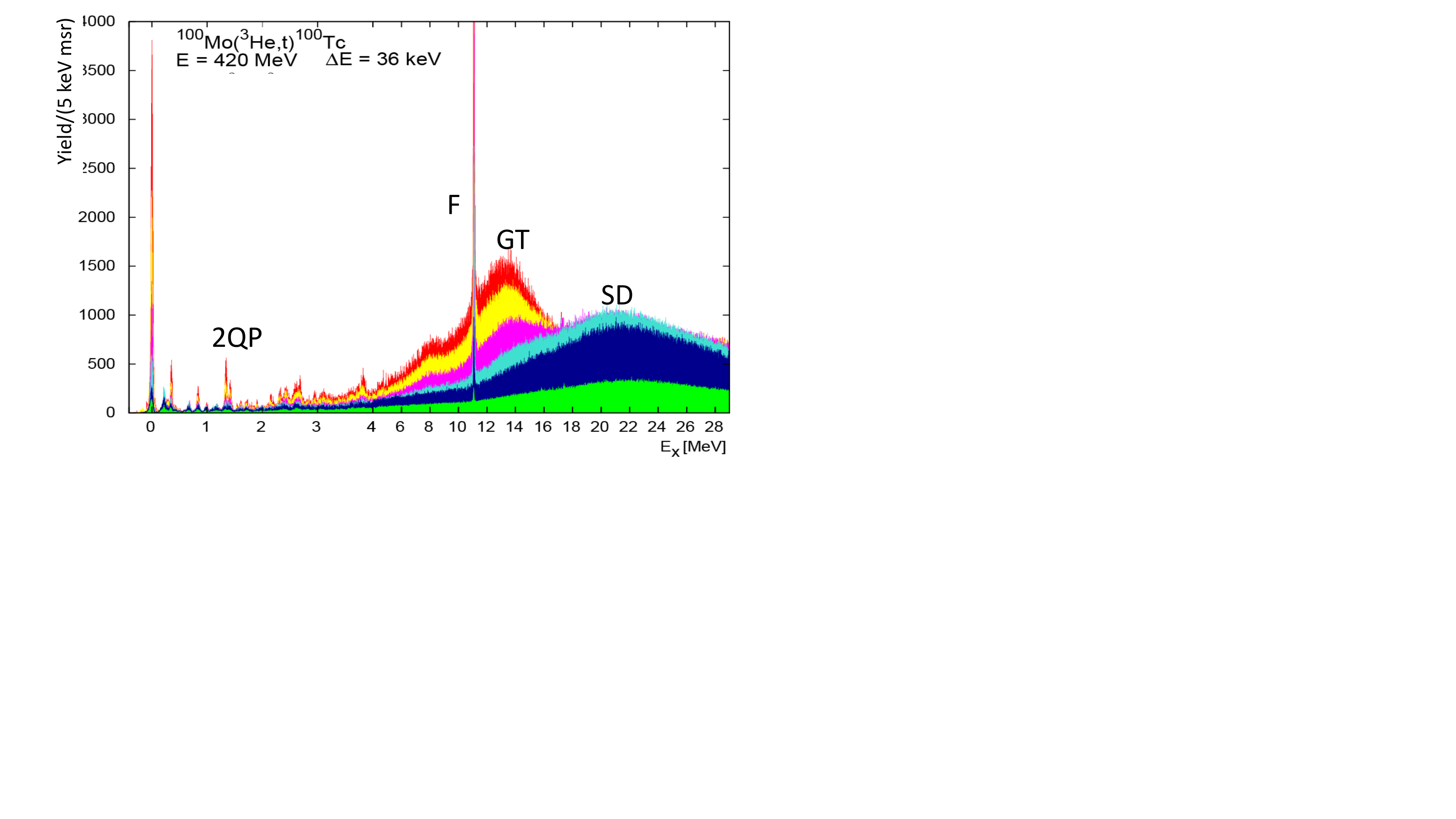}
\vspace{-4.8cm}
\caption{ The energy spectrum of the  ($^3$He,$t$) reaction on $^{100}$Mo \cite{thi12a}. F, GT, and SD are Fermi 0$^+$, GT 1$^+$ and SD 2$^-$ giant resonances, respectively. The energy scale below 4 MeV is enlarged to make the sharp peaks visible. The 1$^-$ and 0$^-$ giant resonances are at the higher energy region of SD. 2QP: Two-quasiparticle states. Yields at the $t$ emission-angles of 0-0.5, 0.5-1, 1-1.5, 1.5-2, 2-2.5, and 2.5-3, each in deg, are shown by red, yellow, pink, light blue, blue, and green, respectively. F and GT transitions are enhanced at the forward angles (red, yellow), while the SD ones at larger angles (blue, green).
\label{figure:fig3}} 
\end{figure}

Recent experimental studies of DBD-NMEs are in \cite{eji19,eji21a}. Among them the charge-exchange ($^3$He,$t$) reactions  provide useful information on  $M^-$(K) with K=GT, SD, and so on for DBD nuclei with $A$=76-136 in a wide excitation-energy and momentum regions, which are relevant to DBD-NMEs \cite{aki97,thi12,thi12b,pup11,thi12a,pup12,eji16a,aki20,eji19a,eji19b}. We discuss the experimental results on the DBD nuclei of $^{76}$Ge,  $^{96}$Zr, $^{100}$Mo, $^{116}$Cd, $^{128}$Te, $^{130}$Te, and $^{136}$Xe.  As an example of the reactions, Fig. 2 shows the $^{100}$Mo spectrum \cite{thi12a}.

The pnQRPA calculations for $M^{0\nu}$
 \cite{jok18} were extended to study  relation of the $M^{0\nu}$  to the 
experimental $E_{\rm GR}(\rm SD)$
and the GT and SD strength distributions. The calculations were made for $g_{\rm A}^{\rm eff}$/$g_{\rm A}$ =0.74 and 0.55  by referring to the experimental data \cite{eji19,eji19a,eji15,eji14} (see (v)). 
 The  modified single-particle energies  so as to reproduce the observed energies for low-lying states were used for all nuclei, except for $^{116}$Cd where the level energies based on the
 Woods-Saxon potential were used.  
 The particle-particle interaction parameter $g_{\rm pp}$ is 
divided in isoscalar and isovector parts in order to recover
 the isospin symmetry (vanishing of the two-neutrino DBD Fermi 
NME) and to reproduce the observed $M^{2\nu}$ \cite{sim13}. 
The $g_{\rm pp}$ 
dependence of $M^{2\nu}$ is visualized, e.g., in \cite{suh07a}
  The particle-hole interaction $g_{\rm ph}$ derived so as to reproduce the experimental  $E_{\rm GR}$(SD) in \cite{eji19}  was used for all multipole transitions except for the GT one, for which the $g'_{\rm ph}$ that fits the experimental $E_{\rm G}$(GT) was used.
 Actually, we first adjust $g_{\rm pp}$ using $g_{\rm ph}$=1.  Then using the adjusted $g_{\rm pp}$, we adjust $g_{\rm ph}$ to fit the GR energy. This is a common way to get the pp and ph interaction parameters to achieve a procedure which is the 
least prone to errors \cite{sim13}. The parameter adjustments for the pnQRPA calculation are well described in \cite{jok18}.

Interesting universal features on the experimental SD data and the  calculated NMEs are found  as given below in (i)-(vi):\\

(i). The GT 1$^+$ strength $B^-({\rm GT})=|M^-({\rm GT})|^2$ and the SD 2$^-$ strength $B^-({\rm SD})=|M^-
({\rm SD})|^2$ are mostly concentrated, respectively, in the GT and SD giant-resonances at the high excitation-energy region in all nuclei (see Fig. 2).  Several sharp peaks in the low (0-6 MeV) energy region are two-quasiparticle states (2QP).
The giant-resonance energies in units of MeV are found to be expressed as 
\begin{equation}
 E_{\rm G}({\rm GT})\approx0.06 A+7.0, ~~E_{\rm G}({\rm SD})\approx0.06A+14.5, 
 \end{equation}
 as shown in Fig. 3 A. They are given also as $E_{\rm G}({\rm GT})\approx0.4 T_{\rm z}$+9 MeV and  $E_{\rm G}({\rm SD})\approx0.4 T_{\rm z}$+16.5 MeV   \cite{eji19}. Note that $A\approx4(N-Z)$+28 for the present nuclei. The $E_{\rm G}(\rm GT)$ and $E_{\rm G}(\rm SD)$ increase gradually as $A$ and $N-Z$ increase.
 This is in accord with the GT and SD giant resonances in other nuclei \cite{hor81}. 
 
 The energies are found to be shifted higher above the GT and SD  2QP states 
by $E_{\rm S}$({\rm GT})$\approx$0.33$(N-Z)$MeV and  $E_{\rm S}$({\rm SD})$\approx$0.3$(N-Z)$MeV. They increase with increase of $N-Z$ and $A$, which reflect the giant-resonance strengths. The SD giant-resonance energies are higher than the GT ones by around 0.9 $\hbar\omega$(harmonic oscillator energy) because of the 1 $\hbar \omega$ jump involved in the SD excitation. 
The giant-resonance energies depend on  $A$ and 
$N-Z$, thus following  the gross properties of nuclear core.
\begin{figure}
\hspace{1cm}
\vspace{-0cm}
\includegraphics[width=0.62\textwidth]{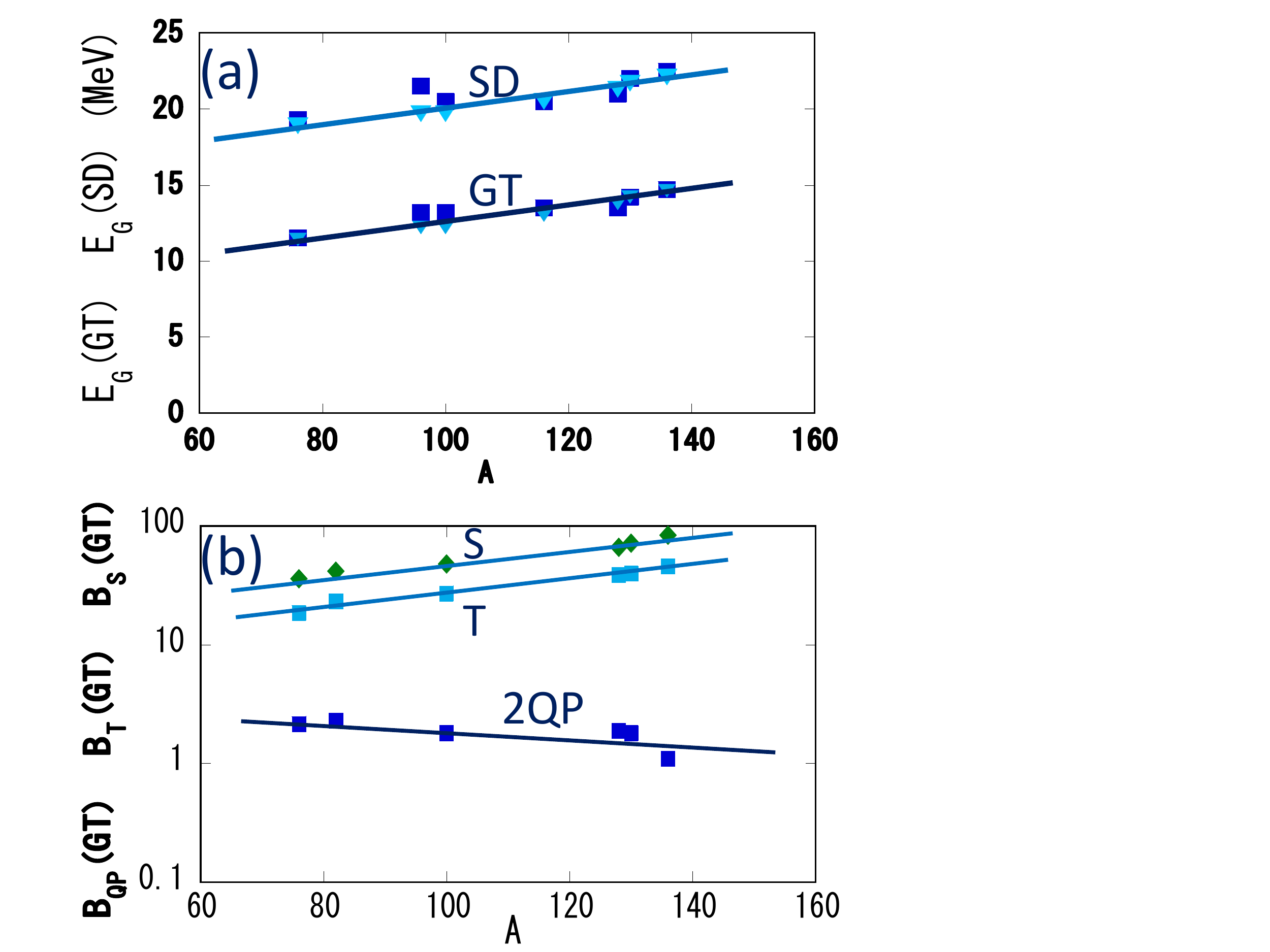}

\vspace{0cm}
\caption{(a): 
Experimental GT 1$^+$ and SD 2$^-$ giant-resonance energies of $E_{\rm G}$(GT) and $E_{\rm G}$(SD). Solid lines: Fits by Eq. (4). Squares: The observed energies.  Inverse triangles: The energies from the experimental  $E_{\rm G}({\rm GT})\approx0.4 T_{\rm z}$+9 MeV and 
$E_{\rm G}({\rm SD})\approx0.4 T_{\rm z}$+16.5 MeV \cite{eji19} as used in the pnQRPA.\\
(b): The GT strengths in logarithmic scale. GT 1$^+$ strength $B^-({\rm GT})$. S:  $B^-_{\rm S}$ (GT) for the nucleon-based sum-rule limit. 
T: $B^-_{\rm T}$(GT)  for the observed total strength up to
 $E_{\rm G}(\rm GT)$+10 MeV.
The strength beyond $E_{\rm G}$ is corrected  for the small contribution from the quasi-free scattering.  2QP: $B^-_{\rm QP}$(GT) for the sum of the observed 2QP strengths up to 6 MeV.  Straight lines are fits to the data.
\label{figure:fig3}} 
\end{figure}

(ii). The observed total GT strengths ($B^-_{\rm T}$({\rm GT}))  
  increase as $A$ and  $N-Z$ increase, and they are around 0.55 of the nucleon-based sum-rule limit of $B^-_{\rm S}$({\rm GT})$\approx$3$(N-Z)$ as shown in Fig. 3 (b) \cite{eji19}. These are consistent with GT strengths in other nuclei \cite{hor81}. The reduction is incorporated by the quenched  axial-vector coupling of $g_{\rm A}^{\rm eff}$/$g_{\rm A}\approx \sqrt{0.55}\approx$0.74. This may stand for such non-nucleonic effects that are not included in the simple sum-rule. The 
2QP GT states are expected in the low excitation-energy region. The summed strength $B^-_{\rm QP}$({\rm GT}) up to 6 MeV is around 10-5 $\%$ of the total strength of $B^-_{\rm T}$(GT), and decrease as $A$ and $N-Z$ increase (see Fig. 3(b)). 

 The GT and SD cross sections for the $(^3$He,$t$) reactions show the maximum at the s- and p-wave $t$ scattering-angles of  $\theta$=0 deg. and around 2 deg., respectively. The GT and SD cross sections there are the same within 10$\%$.
The SD cross section for  the 2QP states below the SD giant resonance  is also nearly same as the GT one below the GT giant resonance. The SD cross sections for the 2QP region are  found to be also of the order of 10-5 $\%$ of the total SD cross sections. Thus the  SD strengths at the giant resonance and the 2QP region show the similar trends as  the GT strengths there. 
The 2QP GT and 2QP SD strengths for the very low excitation-energy (a few MeV) region depend on individual nuclei, reflecting 
the valence nucleon configurations containing single particle  states near the proton and neutron Fermi surfaces of each nucleus.
 

\begin{table}[htb] 
\caption{The pnQRPA NMEs.  $M_{\rm V}^{0\nu}=(g_{\rm V}$/$g_{\rm A}$)$^2M_{\rm F}^{0\nu}$ and $M_{\rm A}^{0\nu}$= ($g_{\rm A}^{\rm eff}$/$g_{\rm A})^2$($M_{\rm GT}^{0\nu}+M_{\rm T}^{0\nu}$) and $M^{0\nu}$ (See  Eq. (2)). $a)$ and $b)$ are the NMEs with $g_{\rm A}^{\rm eff}$/$g_{\rm A}$=0.74 and 0.55, respectively. Note that for $^{100}$Mo the 0g-shell effect on the $M^{2\nu}$  is huge, but looks little on the $M^{0\nu}$.
\label{tab:1}}
\vspace{0.5 cm}
\centering
\begin{tabular}{cccccc}
\hline
$^A$X~& $~M_{\rm V}^{0\nu}$ & $~M_{\rm A}^{0\nu}~^{a)}$ ~  & $~M_A^{0\nu}~^{b)}$ ~& $~M^{0\nu}~^{a)}$~ & $~M^{0\nu}~^{b)}$ \\
\hline
$^{76}$Ge~~ &-1.16 & 2.59 &2.02 & 3.75  & 3.18  \\
$^{96}$Zr~~ &-1.03& 2.12 &1.29& 3.14  & 2.31  \\
$^{100}$Mo~~ &-1.51 & 2.11 &1.81 & 3.62  & 3.32  \\
$^{116}$Cd~~ &-1.01 & 2.03 &1.43 & 3.03  & 2.44  \\
$^{128}$Te~~ &-0.95 & 1.88 &1,29 & 2.82  & 2.24  \\
$^{130}$Te~~ &-0.81 & 1.57 &1.08 & 2.37  & 1.89  \\
$^{136}$Xe~~ &-0.63 & 1.57 &1.05 & 2.19  & 1.68  \\
\hline
\end{tabular}
\end{table}

\begin{figure}[htb]
\hspace{0cm}
\vspace{-0cm}
\includegraphics[width=0.52\textwidth]{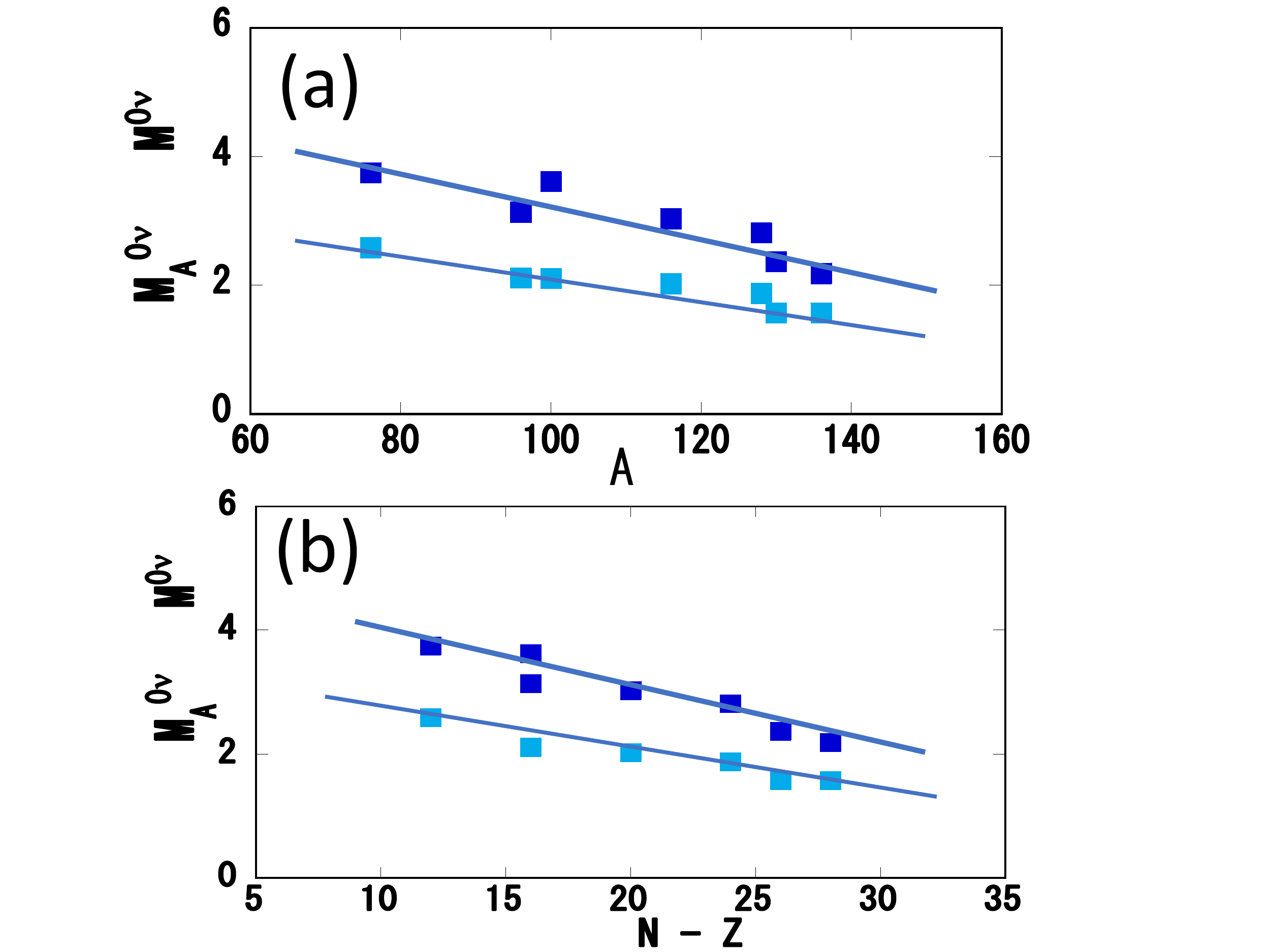}
\vspace{-0.2cm}
\caption{$M^{0\nu}$ (blue squares) and $M^{0\nu}_{\rm A}$(light blue square) with $g_{\rm A}^{\rm eff}$/$g_{\rm A}$=0.74 are plotted against $A$ (panel (a)) and $N-Z$ (panel (b)). Solid lines are fits. See text.
\label{figure:fig4}} 
\end{figure}
(iii). The pnQRPA NMEs of $M^{0\nu}_{\rm A}$=($g_{\rm A}^{\rm eff}$/$g_{\rm A})^2$ [$M_{\rm GT}^{0\nu}+M_T^{0\nu}$] and $M^{0\nu}$ for $g_{\rm A}^{\rm eff}$/$g_{\rm A}$=0.74  (see  Eq. (2)) are shown in Table 1. They are found to decrease gradually as $A$ and $N-Z$ increases, reflecting the gross properties of the nuclear core, as shown in Fig. 4. 
This is in contrast to the SD and GT giant-resonance energies and the strengths, which increase as $A$ and $N-Z$ (see Fig. 3). 
 The  DBD-NMEs are found to behave  like the 2QP GT and the 2QP SD strengths of $B^-_{\rm QP}$(GT) and $B^-_{\rm QP}$(SD) (see Fig. 3 B). 
 
  $M^{2\nu}$, which is associated with the GT NMEs for low-lying 2QP 1$^+$ states, changes more than a factor 10  among the DBD nuclei, depending on the  valence nucleon configurations of states near the Fermi surfaces \cite{eji19,yos13,pir15,eji17}. 
  
(iv). The repulsive $\tau \sigma$ interaction pushes up the GT and SD strengths to the GT and SD giant resonances, and reduces the low-lying GT and SD NMEs with respect to the 2QP NMEs due to the $\tau\sigma$ nuclear polarization effects \cite{eji78,boh75,eji89}. The reduction rate is given by
  1/(1+$\chi$) with $\chi$ being the $\tau\sigma$ susceptibility. Likewise the DBD-NME  $M^{0\nu}$ may  be doubly reduced by the factor (1/$(1+\chi))^2$. Then,  the NME may be expressed as $
M^{0\nu}\approx M_0^{0\nu}$/$(1+\chi)^{2}$
 with $M_0^{0\nu}\approx$6.5 and $\chi\approx$0.025$(N-Z)$ for the present DBD nuclei. As $(N-Z)$ and $A$ increase,  the susceptibility $\chi$
increases,  and $M^{0\nu}$ decreases. The value for $\chi$ is around 0.4 for  DBD nuclei with $N-Z$ around 16, and  $M^{0\nu}\approx$ 0.5 $M^{0\nu}_0$. 

The particle-hole interaction ($g_{\rm ph}$) pushes up the giant resonances in energy, shifting the spin-isospin strengths from the low-lying states to the  giant resonances, and  reduces the  DBD-NME $M^{0\nu}$. The susceptibility $\chi$ in (iv) is proportional to $g_{\rm ph}$ \cite{eji78,eji89}. Thus $\chi$ increases as $g_{\rm ph}$ increases and $M^{0\nu}$ decreases as $g_{\rm ph}$ increases, 10$\%$ increase of $g_{\rm ph}$ leading to around 5 $\%$  decrease of $M^{0\nu}$
in accord with the pnQRPA  calculations  \cite{jok18}.
 
 (v). 
We use for the present DBD NMEs  $(g_{\rm A}^{\rm eff}/g_{\rm A})\approx$ 0.74 and 0.55, which are suggested, respectively, by comparing the observed total GT strength of $B^-_{\rm T}({\rm GT})$ and  the GT and SD strengths for low-lying states with those of the sum rule and  the pnQRPA model \cite{eji19,eji19a,eji19b,eji15,eji14}. The charge-exchange reaction data show the uniform quenching factor over the momentum region of 30-120 MeV/c of the DBD interest \cite{eji19,eji19b}. 
 The coupling of  $g_{\rm A}^{\rm eff}$=1 (i.e.  $g_{\rm A}^{\rm eff}$/$g_{\rm A}$=0.79) is used in \cite{jok18}.  The quenching of the $g_{\rm A}$ for $M^{0\nu}$ is discussed in \cite{eji19,suh17,gys18,eji19a}, and  that for GT and the $M^{2\nu}$ in \cite{eji19,suh17a,coe18,sim18,cor19}.   The similar quenching factor is obtained for the low-momentum GT NMEs from the chiral two-body current \cite{men11}. 
 
 (vi). The NMEs $M_{\rm A}^{0\nu}$= ($g_{\rm A}^{\rm eff}$/$g_{\rm A})^2 [M^{0\nu}_{\rm GT} + M^{0\nu}_{\rm T}]$ with $g_{\rm A}^{\rm eff}$/$g_{\rm A}$=0.55 are smaller than those with  $(g_{\rm A}^{\rm eff}/g_{\rm A})$= 0.74 by factors around 0.7. The reduction is less severe than 0.62 for the ratio of the $(g_{\rm A}^{\rm eff}$/$g_{\rm A})^2 $. 
The difference is due to the dependence of the 
computed NME $M^{2\nu}$ on the ratio 
$g_{\rm A}^{\rm eff}/g_{\rm A}$ and hence on $g_{\rm pp}$.
 As 
$g_{\rm A}^{\rm eff}/g_{\rm A}$ decreases the magnitude
 of the
 experimental NME, derived from the experimental half-life, 
increases and hence also the magnitude of the computed NME has 
to increase.
 Since the magnitude of the computed $M^{2\nu}$ 
increases with decreasing value of $g_{\rm A}^{\rm eff}/
g_{\rm A}$ \cite{suh07a} the value of $g_{\rm pp}$ has to decrease. This, 
in turn, means that the NMEs $M^{0\nu}$ are computed with 
different values of 
$g_{\rm pp}$ for different values of the ratio 
$g_{\rm A}^{\rm eff}/g_{\rm A}$ (larger ratio means larger
 value of $g_{\rm pp}$).
  $M^{0\nu}$ 
behaves in a similar way in terms of $g_{\rm A}^{\rm eff}/g_{\rm A}$ as 
$M^{2\nu}$ (less sensitively, though, since the higher
multipoles are less dependent on the ratio 
$g_{\rm A}^{\rm eff}/g_{\rm A}$ than the $1^+$ multipole). 
The computed NMEs $M^{0\nu}$
 with $g_{\rm A}^{\rm eff}/g_{\rm A}=0.55$ are smaller by
 factors around 0.8 than those with
$g_{\rm A}^{\rm eff}/g_{\rm A}=0.74$.\\

  Let us discuss the impact of the present findings on DBD studies. 
On the bases of the above discussions, we use $g_{\rm A}^{\rm eff}/g_A\approx$0.65$\pm{0.1}$.
 $M^{0\nu}$ with  this $g_{\rm A}^{\rm eff}$/$g_{\rm A}$ region corresponds approximately to the $\pm{10}\%$ region around   
\begin{equation}
M^{0\nu}\approx 5.2-0.023 A, ~~~
M^{0\nu}\approx 4.2-0.08 (N-Z). 
\end{equation}
The DBD NMEs for $A$=76-136 are considered to be around 3-2.  Since they  do not depend much on individual nuclei, selection of the DBD nucleus for the experiment may be made from  experimental requirements such as the availability of  ton-scale DBD isotopes, the large $Q$ value, the low-background and  the high energy-resolution.


  
  The present analyses show that the pnQRPA $M^{0\nu}$  is closely related to the SD giant resonance and the distribution of the strength $B^-$(K)= $|M^-(\rm K)|^2$ with K=GT, SD in the intermediate nucleus. Various  nuclear models are being used for calculating  $M^{0\nu}$. Then, it is interesting to compare the calculated strength distribution of $B^-(\rm K)$ in $^A_{Z+1}$X$_2$ by using the model wave function for $^A_Z$X$_1$ with the observed giant resonance and the strength. 

 Actually, the SD ($J^{\pi}$=2$^-$) giant resonance is accompanied at the higher energy side by the resonances with $J^{\pi}$=1$^-$, 0$^-$ and the quasi-free scattering. 
Further experimental studies for their strengths and also for higher multipole strengths with $J^{\pi}$=3$^+$ and $J^{\pi}$=4$^-$ are interesting. The low and high multipole giant resonances are studied in the pnQRPA \cite{jok17}.

OMC(ordinary muon capture) of  ($\mu,\nu_{\mu}$) is used to study $M^+$(K) \cite{eji19,has18}. The large strength is found  in the giant resonances \cite{has18,jok19}. The quenching of $g_{\rm A}$ in the muon strengths is under discussions \cite{jok19a,sim20}. Actually, $\tau^{\pm}$ strengths of  $B^{\pm}$(K)
are also studied by using various
nuclear reactions \cite{eji19,eji20,eji21a}.

The $\tau^-$  strengths in the intermediate nucleus $^A_{Z+1}$X$_2$ are mostly in the giant resonances, and little strengths at the  2QP and ground states. Likewise, the DBD ($\tau^- \tau^-$) strengths are considered to be mostly in the double giant resonances, and little DBD strengths at the 2QP and ground states (Fig. 1). 
Double charge-exchange reactions on $^{A}_{Z}$X$_1$ are used  to study  the DBD strengths in  $^A_{Z+2}$X$_3$ \cite{eji19,cap15}.  The double GT and SD giant-resonance energies 
measured from the ground state of $^A_Z$X$_1$ are around 
 $E_{\rm G'}$(DGT)=26-32 MeV and $E_{\rm G'}$(DSD)=42-48 MeV for DBD nuclei with $A$=76-136. 
The RCNP ($^{11}$B,$^{11}$Li) data \cite{tak17} show that most of the strengths are at the high excitation-energy region. The double GT strengths for the ground states are shown to have a positive correlation with the DGT-state DBD-NMEs \cite{shi18}. 
  
  The delta isobar ($\Delta$) is strongly excited by the quark $\sigma \tau$ flip  in a nucleon to form the N$^{-1}\Delta$ giant resonance.  
 Then, the  axial-vector $M^{\pm}$(K) and the axial-vector  $M_{\alpha}^{0\nu}$ with $\alpha$=GT and T are reduced with respect to the model NMEs without the $\Delta$ isobar effect, which
 is incorporated by using the effective coupling of $g_{\rm A}^{\rm eff}$ 
  \cite{boh81,eji82,cat02,eji19,eji21b}. Studies of the N$^{-1}\Delta$ effects are interesting. 
  
 In conclusion, the present work shows for the first time (i) a clear relation between the experimental SD strength distribution in the intermediate nucleus and the DBD NME in the pnQRPA formalism based on the SD data, and (ii) a simple expression of  $M^{0\nu}$  as a function of $A$ on the basis of the pnQRPA with the $g_{\rm A}^{\rm eff}$/$g_{\rm A}$ derived experimentally. Thus such experimental inputs are useful for pinning down the values for the DBD NMEs.


\begin{thebibliography}{62}
\providecommand{\natexlab}[1]{#1}
\providecommand{\url}[1]{\texttt{#1}}
\expandafter\ifx\csname urlstyle\endcsname\relax
  \providecommand{\doi}[1]{doi: #1}\else
  \providecommand{\doi}{doi: \begingroup \urlstyle{rm}\Url}\fi

\bibitem[Ba et~al.(2016)Ba, Kiros, and Hinton]{layer_norm}
Lei~Jimmy Ba, Jamie~Ryan Kiros, and Geoffrey~E. Hinton.
\newblock Layer normalization.
\newblock \emph{arXiv preprint}, arXiv:1607.06450, 2016.

\bibitem[Battaglia et~al.(2018)Battaglia, Hamrick, Bapst, Sanchez{-}Gonzalez,
  Zambaldi, Malinowski, Tacchetti, Raposo, Santoro, Faulkner,
  G{\"{u}}l{\c{c}}ehre, Song, Ballard, Gilmer, Dahl, Vaswani, Allen, Nash,
  Langston, Dyer, Heess, Wierstra, Kohli, Botvinick, Vinyals, Li, and
  Pascanu]{peter_survey}
Peter~W. Battaglia, Jessica~B. Hamrick, Victor Bapst, Alvaro
  Sanchez{-}Gonzalez, Vin{\'{\i}}cius~Flores Zambaldi, Mateusz Malinowski,
  Andrea Tacchetti, David Raposo, Adam Santoro, Ryan Faulkner, {\c{C}}aglar
  G{\"{u}}l{\c{c}}ehre, H.~Francis Song, Andrew~J. Ballard, Justin Gilmer,
  George~E. Dahl, Ashish Vaswani, Kelsey~R. Allen, Charles Nash, Victoria
  Langston, Chris Dyer, Nicolas Heess, Daan Wierstra, Pushmeet Kohli, Matthew
  Botvinick, Oriol Vinyals, Yujia Li, and Razvan Pascanu.
\newblock Relational inductive biases, deep learning, and graph networks.
\newblock \emph{arXiv preprint arXiv:1806.01261}, 2018.

\bibitem[Chang et~al.(2020)Chang, Rong, Xu, Huang, Zhang, Cui, Zhu, and
  Huang]{restrict_blk_adv}
Heng Chang, Yu~Rong, Tingyang Xu, Wenbing Huang, Honglei Zhang, Peng Cui, Wenwu
  Zhu, and Junzhou Huang.
\newblock A restricted black-box adversarial framework towards attacking graph
  embedding models.
\newblock In \emph{The Thirty-Fourth {AAAI} Conference on Artificial
  Intelligence}, pp.\  3389--3396, 2020.

\bibitem[Dai et~al.(2018)Dai, Li, Tian, Huang, Wang, Zhu, and Song]{rls2v}
Hanjun Dai, Hui Li, Tian Tian, Xin Huang, Lin Wang, Jun Zhu, and Le~Song.
\newblock Adversarial attack on graph structured data.
\newblock In \emph{Proceedings of the 35th International Conference on Machine
  Learning}, pp.\  1123--1132, 2018.

\bibitem[Entezari et~al.(2020)Entezari, Al{-}Sayouri, Darvishzadeh, and
  Papalexakis]{gnn_svd}
Negin Entezari, Saba~A. Al{-}Sayouri, Amirali Darvishzadeh, and Evangelos~E.
  Papalexakis.
\newblock All you need is low (rank): Defending against adversarial attacks on
  graphs.
\newblock In \emph{The Thirteenth {ACM} International Conference on Web Search
  and Data Mining}, pp.\  169--177, 2020.

\bibitem[Feng et~al.(2021)Feng, He, Tang, and Chua]{dyadv}
Fuli Feng, Xiangnan He, Jie Tang, and Tat{-}Seng Chua.
\newblock Graph adversarial training: Dynamically regularizing based on graph
  structure.
\newblock \emph{IEEE Transactions on Knowledge and Data Engineering},
  33\penalty0 (6):\penalty0 2493--2504, 2021.

\bibitem[Fey \& Lenssen(2019)Fey and Lenssen]{pytorch_geometric}
Matthias Fey and Jan~E. Lenssen.
\newblock Fast graph representation learning with {PyTorch Geometric}.
\newblock In \emph{ICLR Workshop on Representation Learning on Graphs and
  Manifolds}, 2019.

\bibitem[Finn et~al.(2017)Finn, Abbeel, and Levine]{maml}
Chelsea Finn, Pieter Abbeel, and Sergey Levine.
\newblock Model-agnostic meta-learning for fast adaptation of deep networks.
\newblock In \emph{Proceedings of the 34th International Conference on Machine
  Learning}, pp.\  1126--1135, 2017.

\bibitem[Giles et~al.(1998)Giles, Bollacker, and Lawrence]{citeseer}
C.~Lee Giles, Kurt~D. Bollacker, and Steve Lawrence.
\newblock Citeseer: An automatic citation indexing system.
\newblock In \emph{Proceedings of the 3rd {ACM} International Conference on
  Digital Libraries}, pp.\  89--98, 1998.

\bibitem[Goodfellow et~al.(2016)Goodfellow, Bengio, and Courville]{dl_book}
Ian Goodfellow, Yoshua Bengio, and Aaron Courville.
\newblock \emph{Deep Learning}.
\newblock 2016.
\newblock \url{http://www.deeplearningbook.org}.

\bibitem[Goodfellow et~al.(2015)Goodfellow, Shlens, and Szegedy]{fgsm}
Ian~J. Goodfellow, Jonathon Shlens, and Christian Szegedy.
\newblock Explaining and harnessing adversarial examples.
\newblock In \emph{International Conference on Learning Representations}, 2015.

\bibitem[Hamilton et~al.(2017{\natexlab{a}})Hamilton, Ying, and
  Leskovec]{jure_grlsurvey}
William~L. Hamilton, Rex Ying, and Jure Leskovec.
\newblock Representation learning on graphs: Methods and applications.
\newblock \emph{{IEEE} Data Engineering Bulletin}, 40\penalty0 (3):\penalty0
  52--74, 2017{\natexlab{a}}.

\bibitem[Hamilton et~al.(2017{\natexlab{b}})Hamilton, Ying, and Leskovec]{sage}
William~L. Hamilton, Zhitao Ying, and Jure Leskovec.
\newblock Inductive representation learning on large graphs.
\newblock In \emph{Advances in Neural Information Processing Systems}, pp.\
  1024--1034, 2017{\natexlab{b}}.

\bibitem[Han et~al.(2018)Han, Yao, Yu, Niu, Xu, Hu, Tsang, and
  Sugiyama]{coteaching}
Bo~Han, Quanming Yao, Xingrui Yu, Gang Niu, Miao Xu, Weihua Hu, Ivor Tsang, and
  Masashi Sugiyama.
\newblock Co-teaching: Robust training of deep neural networks with extremely
  noisy labels.
\newblock In \emph{Advances in Neural Information Processing Systems}, 2018.

\bibitem[Han et~al.(2020{\natexlab{a}})Han, Niu, Yu, Yao, Xu, Tsang, and
  Sugiyama]{sigua}
Bo~Han, Gang Niu, Xingrui Yu, Quanming Yao, Miao Xu, Ivor Tsang, and Masashi
  Sugiyama.
\newblock {SIGUA}: Forgetting may make learning with noisy labels more robust.
\newblock In \emph{Proceedings of the 37th International Conference on Machine
  Learning}, pp.\  4006--4016, 2020{\natexlab{a}}.

\bibitem[Han et~al.(2020{\natexlab{b}})Han, Yao, Liu, Niu, Tsang, Kwok, and
  Sugiyama]{label_noise_survey}
Bo~Han, Quanming Yao, Tongliang Liu, Gang Niu, Ivor~W. Tsang, James~T. Kwok,
  and Masashi Sugiyama.
\newblock A survey of label-noise representation learning: Past, present and
  future.
\newblock \emph{arXiv preprint}, arXiv:2011.04406, 2020{\natexlab{b}}.

\bibitem[Hornik et~al.(1989)Hornik, Stinchcombe, and White]{mlp_approx}
Kurt Hornik, Maxwell~B. Stinchcombe, and Halbert White.
\newblock Multilayer feedforward networks are universal approximators.
\newblock \emph{Neural Networks}, 2\penalty0 (5):\penalty0 359--366, 1989.

\bibitem[Hou et~al.(2020)Hou, Zhang, Cheng, Ma, Ma, Chen, and
  Yang]{Hou2020Measuring}
Yifan Hou, Jian Zhang, James Cheng, Kaili Ma, Richard T.~B. Ma, Hongzhi Chen,
  and Ming-Chang Yang.
\newblock Measuring and improving the use of graph information in graph neural
  networks.
\newblock In \emph{International Conference on Learning Representations}, 2020.

\bibitem[Hu et~al.(2020)Hu, Fey, Zitnik, Dong, Ren, Liu, Catasta, and
  Leskovec]{ogb}
Weihua Hu, Matthias Fey, Marinka Zitnik, Yuxiao Dong, Hongyu Ren, Bowen Liu,
  Michele Catasta, and Jure Leskovec.
\newblock Open graph benchmark: Datasets for machine learning on graphs.
\newblock In \emph{Advances in Neural Information Processing Systems}, 2020.

\bibitem[Jin \& Zhang(2021)Jin and Zhang]{laten_adv}
Hongwei Jin and Xinhua Zhang.
\newblock Robust training of graph convolutional networks via latent
  perturbation.
\newblock In \emph{Machine Learning and Knowledge Discovery in Databases}, pp.\
   394--411, 2021.

\bibitem[Jin et~al.(2020)Jin, Ma, Liu, Tang, Wang, and Tang]{prognn}
Wei Jin, Yao Ma, Xiaorui Liu, Xianfeng Tang, Suhang Wang, and Jiliang Tang.
\newblock Graph structure learning for robust graph neural networks.
\newblock In \emph{The 26th {ACM} {SIGKDD} Conference on Knowledge Discovery
  and Data Mining}, pp.\  66--74, 2020.

\bibitem[Jin et~al.(2021)Jin, Li, Xu, Wang, Ji, Aggarwal, and Tang]{deeprobust}
Wei Jin, Yaxing Li, Han Xu, Yiqi Wang, Shuiwang Ji, Charu Aggarwal, and Jiliang
  Tang.
\newblock Adversarial attacks and defenses on graphs.
\newblock \emph{SIGKDD Explorations Newsletter}, 22\penalty0 (2):\penalty0
  19–34, 2021.

\bibitem[Kingma \& Ba(2015)Kingma and Ba]{adam}
Diederik~P. Kingma and Jimmy Ba.
\newblock Adam: {A} method for stochastic optimization.
\newblock In \emph{International Conference on Learning Representations}, 2015.

\bibitem[Kipf \& Welling(2017)Kipf and Welling]{gcn}
Thomas~N. Kipf and Max Welling.
\newblock Semi-supervised classification with graph convolutional networks.
\newblock In \emph{International Conference on Learning Representations}, 2017.

\bibitem[Klicpera et~al.(2019)Klicpera, Bojchevski, and
  Günnemann]{klicpera2018combining}
Johannes Klicpera, Aleksandar Bojchevski, and Stephan Günnemann.
\newblock Combining neural networks with personalized pagerank for
  classification on graphs.
\newblock In \emph{International Conference on Learning Representations}, 2019.

\bibitem[Kong et~al.(2020)Kong, Li, Ding, Wu, Zhu, Ghanem, Taylor, and
  Goldstein]{flag}
Kezhi Kong, Guohao Li, Mucong Ding, Zuxuan Wu, Chen Zhu, Bernard Ghanem, Gavin
  Taylor, and Tom Goldstein.
\newblock {FLAG:} adversarial data augmentation for graph neural networks.
\newblock \emph{arXiv preprint}, arXiv:2010.09891, 2020.

\bibitem[Liu \& Tao(2016)Liu and Tao]{label_noise_reweight}
Tongliang Liu and Dacheng Tao.
\newblock Classification with noisy labels by importance reweighting.
\newblock \emph{IEEE Transactions on Pattern Analysis \& Machine Intelligence},
  38\penalty0 (03):\penalty0 447--461, 2016.

\bibitem[London \& Getoor(2014)London and Getoor]{homophily2_collective}
Ben London and Lise Getoor.
\newblock Collective classification of network data.
\newblock In \emph{Data Classification: Algorithms and Applications}, pp.\
  399--416. 2014.

\bibitem[Ma et~al.(2022)Ma, Liu, Shah, and Tang]{is_homophily_necessity}
Yao Ma, Xiaorui Liu, Neil Shah, and Jiliang Tang.
\newblock Is homophily a necessity for graph neural networks?
\newblock In \emph{International Conference on Learning Representations}, 2022.

\bibitem[Madry et~al.(2018)Madry, Makelov, Schmidt, Tsipras, and Vladu]{pgd}
Aleksander Madry, Aleksandar Makelov, Ludwig Schmidt, Dimitris Tsipras, and
  Adrian Vladu.
\newblock Towards deep learning models resistant to adversarial attacks.
\newblock In \emph{International Conference on Learning Representations}, 2018.

\bibitem[McAuley et~al.(2015)McAuley, Targett, Shi, and van~den
  Hengel]{computers_photo}
Julian~J. McAuley, Christopher Targett, Qinfeng Shi, and Anton van~den Hengel.
\newblock Image-based recommendations on styles and substitutes.
\newblock In \emph{Proceedings of the 38th International {ACM} {SIGIR}
  Conference on Research and Development in Information Retrieval}, pp.\
  43--52, 2015.

\bibitem[McPherson et~al.(2001)McPherson, Smith-Lovin, and
  Cook]{homophily1_birds}
Miller McPherson, Lynn Smith-Lovin, and James~M Cook.
\newblock Birds of a feather: Homophily in social networks.
\newblock \emph{Annual Review of Sociology}, 27\penalty0 (1):\penalty0
  415--444, 2001.

\bibitem[Paszke et~al.(2019)Paszke, Gross, Massa, Lerer, Bradbury, Chanan,
  Killeen, Lin, Gimelshein, Antiga, Desmaison, Kopf, Yang, DeVito, Raison,
  Tejani, Chilamkurthy, Steiner, Fang, Bai, and Chintala]{pytorch}
Adam Paszke, Sam Gross, Francisco Massa, Adam Lerer, James Bradbury, Gregory
  Chanan, Trevor Killeen, Zeming Lin, Natalia Gimelshein, Luca Antiga, Alban
  Desmaison, Andreas Kopf, Edward Yang, Zachary DeVito, Martin Raison, Alykhan
  Tejani, Sasank Chilamkurthy, Benoit Steiner, Lu~Fang, Junjie Bai, and Soumith
  Chintala.
\newblock Pytorch: An imperative style, high-performance deep learning library.
\newblock In \emph{Advances in Neural Information Processing Systems}, pp.\
  8024--8035, 2019.

\bibitem[Pei et~al.(2020)Pei, Wei, Chang, Lei, and Yang]{geom_gcn}
Hongbin Pei, Bingzhe Wei, Kevin~Chen{-}Chuan Chang, Yu~Lei, and Bo~Yang.
\newblock Geom-gcn: Geometric graph convolutional networks.
\newblock In \emph{International Conference on Learning Representations}, 2020.

\bibitem[Rong et~al.(2020)Rong, Huang, Xu, and Huang]{dropedge}
Yu~Rong, Wenbing Huang, Tingyang Xu, and Junzhou Huang.
\newblock Dropedge: Towards deep graph convolutional networks on node
  classification.
\newblock In \emph{International Conference on Learning Representations}, 2020.

\bibitem[Srivastava et~al.(2014)Srivastava, Hinton, Krizhevsky, Sutskever, and
  Salakhutdinov]{dropout}
Nitish Srivastava, Geoffrey~E. Hinton, Alex Krizhevsky, Ilya Sutskever, and
  Ruslan Salakhutdinov.
\newblock Dropout: a simple way to prevent neural networks from overfitting.
\newblock \emph{Journal of Machine Learning Research}, 15\penalty0
  (1):\penalty0 1929--1958, 2014.

\bibitem[Sun et~al.(2018)Sun, Dou, Yang, Wang, Yu, and Li]{sun_gadv_survey}
Lichao Sun, Yingtong Dou, Carl Yang, Ji~Wang, Philip~S. Yu, and Bo~Li.
\newblock Adversarial attack and defense on graph data: {A} survey.
\newblock \emph{arXiv preprint}, arXiv:1812.10528, 2018.

\bibitem[Sun et~al.(2020)Sun, Wang, Tang, Hsieh, and Honavar]{nipa}
Yiwei Sun, Suhang Wang, Xianfeng Tang, Tsung{-}Yu Hsieh, and Vasant~G. Honavar.
\newblock Adversarial attacks on graph neural networks via node injections: {A}
  hierarchical reinforcement learning approach.
\newblock In \emph{The Web Conference 2020}, pp.\  673--683, 2020.

\bibitem[Szegedy et~al.(2014)Szegedy, Zaremba, Sutskever, Bruna, Erhan,
  Goodfellow, and Fergus]{intriguing}
Christian Szegedy, Wojciech Zaremba, Ilya Sutskever, Joan Bruna, Dumitru Erhan,
  Ian~J. Goodfellow, and Rob Fergus.
\newblock Intriguing properties of neural networks.
\newblock In \emph{International Conference on Learning Representations}, 2014.

\bibitem[Tang et~al.(2008)Tang, Zhang, Yao, Li, Zhang, and Su]{aminer}
Jie Tang, Jing Zhang, Limin Yao, Juanzi Li, Li~Zhang, and Zhong Su.
\newblock Arnetminer: extraction and mining of academic social networks.
\newblock In \emph{Proceedings of the 14th {ACM} {SIGKDD} International
  Conference on Knowledge Discovery and Data Mining}, pp.\  990--998, 2008.

\bibitem[Veli{\v{c}}kovi{\'{c}} et~al.(2018)Veli{\v{c}}kovi{\'{c}}, Cucurull,
  Casanova, Romero, Li{\`{o}}, and Bengio]{gat}
Petar Veli{\v{c}}kovi{\'{c}}, Guillem Cucurull, Arantxa Casanova, Adriana
  Romero, Pietro Li{\`{o}}, and Yoshua Bengio.
\newblock {Graph Attention Networks}.
\newblock In \emph{International Conference on Learning Representations}, 2018.

\bibitem[Wang et~al.(2020)Wang, Luo, Suya, Li, Yang, and Zheng]{afgsm}
Jihong Wang, Minnan Luo, Fnu Suya, Jundong Li, Zijiang Yang, and Qinghua Zheng.
\newblock Scalable attack on graph data by injecting vicious nodes.
\newblock In \emph{Proceedings of the European Conference on Machine Learning
  and Principles and Practice of Knowledge Discovery in Databases}, 2020.

\bibitem[Wang et~al.(2018)Wang, Cheng, Eaton, Hsieh, and Wu]{fakenodes}
Xiaoyun Wang, Minhao Cheng, Joe Eaton, Cho{-}Jui Hsieh, and Shyhtsun~Felix Wu.
\newblock Attack graph convolutional networks by adding fake nodes.
\newblock \emph{arXiv preprint}, arXiv:1810.10751, 2018.

\bibitem[Wu et~al.(2019)Wu, Wang, Tyshetskiy, Docherty, Lu, and
  Zhu]{advsample_deepinsights}
Huijun Wu, Chen Wang, Yuriy Tyshetskiy, Andrew Docherty, Kai Lu, and Liming
  Zhu.
\newblock Adversarial examples for graph data: Deep insights into attack and
  defense.
\newblock In \emph{Proceedings of the Twenty-Eighth International Joint
  Conference on Artificial Intelligence}, pp.\  4816--4823, 2019.

\bibitem[Wu et~al.(2021)Wu, Pan, Chen, Long, Zhang, and Yu]{wu_survey}
Zonghan Wu, Shirui Pan, Fengwen Chen, Guodong Long, Chengqi Zhang, and
  Philip~S. Yu.
\newblock A comprehensive survey on graph neural networks.
\newblock \emph{IEEE Transactions on Neural Networks and Learning Systems},
  32\penalty0 (1):\penalty0 4--24, 2021.

\bibitem[Xu et~al.(2019{\natexlab{a}})Xu, Chen, Liu, Chen, Weng, Hong, and
  Lin]{topo_atk}
Kaidi Xu, Hongge Chen, Sijia Liu, Pin{-}Yu Chen, Tsui{-}Wei Weng, Mingyi Hong,
  and Xue Lin.
\newblock Topology attack and defense for graph neural networks: An
  optimization perspective.
\newblock In \emph{Proceedings of the Twenty-Eighth International Joint
  Conference on Artificial Intelligence}, pp.\  3961--3967, 2019{\natexlab{a}}.

\bibitem[Xu et~al.(2018)Xu, Li, Tian, Sonobe, Kawarabayashi, and
  Jegelka]{jknet}
Keyulu Xu, Chengtao Li, Yonglong Tian, Tomohiro Sonobe, Ken{-}ichi
  Kawarabayashi, and Stefanie Jegelka.
\newblock Representation learning on graphs with jumping knowledge networks.
\newblock In \emph{Proceedings of the 35th International Conference on Machine
  Learning}, pp.\  5449--5458, 2018.

\bibitem[Xu et~al.(2019{\natexlab{b}})Xu, Hu, Leskovec, and Jegelka]{gin}
Keyulu Xu, Weihua Hu, Jure Leskovec, and Stefanie Jegelka.
\newblock How powerful are graph neural networks?
\newblock In \emph{International Conference on Learning Representations},
  2019{\natexlab{b}}.

\bibitem[Yang et~al.(2021{\natexlab{a}})Yang, Ma, and Cheng]{p-reg}
Han Yang, Kaili Ma, and James Cheng.
\newblock Rethinking graph regularization for graph neural networks.
\newblock In \emph{Thirty-Fifth {AAAI} Conference on Artificial Intelligence},
  pp.\  4573--4581, 2021{\natexlab{a}}.

\bibitem[Yang et~al.(2021{\natexlab{b}})Yang, Yan, Dai, Chen, and
  Cheng]{self-enhance}
Han Yang, Xiao Yan, Xinyan Dai, Yongqiang Chen, and James Cheng.
\newblock Self-enhanced gnn: Improving graph neural networks using model
  outputs.
\newblock In \emph{The International Joint Conference on Neural Networks}, pp.\
   1--8, 2021{\natexlab{b}}.

\bibitem[Yang et~al.(2016)Yang, Cohen, and Salakhutdinov]{cora}
Zhilin Yang, William~W. Cohen, and Ruslan Salakhutdinov.
\newblock Revisiting semi-supervised learning with graph embeddings.
\newblock In \emph{Proceedings of the 33nd International Conference on Machine
  Learning}, pp.\  40--48, 2016.

\bibitem[Zeng et~al.(2020)Zeng, Zhou, Srivastava, Kannan, and Prasanna]{saint}
Hanqing Zeng, Hongkuan Zhou, Ajitesh Srivastava, Rajgopal Kannan, and Viktor~K.
  Prasanna.
\newblock Graphsaint: Graph sampling based inductive learning method.
\newblock In \emph{International Conference on Learning Representations}, 2020.

\bibitem[Zhang et~al.(2017)Zhang, Bengio, Hardt, Recht, and
  Vinyals]{memorization}
Chiyuan Zhang, Samy Bengio, Moritz Hardt, Benjamin Recht, and Oriol Vinyals.
\newblock Understanding deep learning requires rethinking generalization.
\newblock In \emph{International Conference on Learning Representations}, 2017.

\bibitem[Zhang \& Zitnik(2020)Zhang and Zitnik]{gnnguard}
Xiang Zhang and Marinka Zitnik.
\newblock Gnnguard: Defending graph neural networks against adversarial
  attacks.
\newblock In \emph{Advances in Neural Information Processing Systems}, 2020.

\bibitem[Zhao et~al.(2021)Zhao, Liu, Neves, Woodford, Jiang, and Shah]{aug_gnn}
Tong Zhao, Yozen Liu, Leonardo Neves, Oliver~J. Woodford, Meng Jiang, and Neil
  Shah.
\newblock Data augmentation for graph neural networks.
\newblock In \emph{Thirty-Fifth {AAAI} Conference on Artificial Intelligence},
  pp.\  11015--11023, 2021.

\bibitem[Zheng et~al.(2021)Zheng, Zou, Dong, Cen, Yin, Xu, Yang, and Tang]{grb}
Qinkai Zheng, Xu~Zou, Yuxiao Dong, Yukuo Cen, Da~Yin, Jiarong Xu, Yang Yang,
  and Jie Tang.
\newblock Graph robustness benchmark: Benchmarking the adversarial robustness
  of graph machine learning.
\newblock \emph{Neural Information Processing Systems Track on Datasets and
  Benchmarks}, 2021.

\bibitem[Zhou et~al.(2020)Zhou, Cui, Hu, Zhang, Yang, Liu, Wang, Li, and
  Sun]{zhou_survey}
Jie Zhou, Ganqu Cui, Shengding Hu, Zhengyan Zhang, Cheng Yang, Zhiyuan Liu,
  Lifeng Wang, Changcheng Li, and Maosong Sun.
\newblock Graph neural networks: {A} review of methods and applications.
\newblock \emph{{AI} Open}, 1:\penalty0 57--81, 2020.

\bibitem[Zhu et~al.(2019)Zhu, Zhang, Cui, and Zhu]{robustgcn}
Dingyuan Zhu, Ziwei Zhang, Peng Cui, and Wenwu Zhu.
\newblock Robust graph convolutional networks against adversarial attacks.
\newblock In \emph{Proceedings of the 25th {ACM} {SIGKDD} International
  Conference on Knowledge Discovery {\&} Data Mining}, pp.\  1399--1407, 2019.

\bibitem[Zhu et~al.(2020)Zhu, Yan, Zhao, Heimann, Akoglu, and
  Koutra]{beyond_homophily}
Jiong Zhu, Yujun Yan, Lingxiao Zhao, Mark Heimann, Leman Akoglu, and Danai
  Koutra.
\newblock Beyond homophily in graph neural networks: Current limitations and
  effective designs.
\newblock In \emph{Advances in Neural Information Processing Systems}, 2020.

\bibitem[Zou et~al.(2021)Zou, Zheng, Dong, Guan, Kharlamov, Lu, and
  Tang]{tdgia}
Xu~Zou, Qinkai Zheng, Yuxiao Dong, Xinyu Guan, Evgeny Kharlamov, Jialiang Lu,
  and Jie Tang.
\newblock Tdgia: Effective injection attacks on graph neural networks.
\newblock In \emph{Proceedings of the 27th ACM SIGKDD Conference on Knowledge
  Discovery \& Data Mining}, pp.\  2461–2471, 2021.

\bibitem[Z{\"u}gner et~al.(2018)Z{\"u}gner, Akbarnejad, and
  G{\"u}nnemann]{nettack}
Daniel Z{\"u}gner, Amir Akbarnejad, and Stephan G{\"u}nnemann.
\newblock Adversarial attacks on neural networks for graph data.
\newblock In \emph{Proceedings of the 24th {ACM} {SIGKDD} International
  Conference on Knowledge Discovery {\&} Data Mining}, pp.\  2847--2856, 2018.

\bibitem[Zügner \& Günnemann(2019)Zügner and Günnemann]{metattack}
Daniel Zügner and Stephan Günnemann.
\newblock Adversarial attacks on graph neural networks via meta learning.
\newblock In \emph{7th International Conference on Learning Representations},
  2019.

\end{thebibliography}


\begin{thebibliography}{00}


\bibitem {eji05} 
 H. Ejiri, 
 J. Phys. Soc. Jpn. {\bf 74}, 2101(2005). 
  
\bibitem{avi08}  
F. Avignone, S.  Elliott, and J.  Engel, 
 Rev. Mod. Phys. {\bf 80}, 481 (2008).


\bibitem {ver12} 
J. Vergados, H.  Ejiri, and F.  {\v S}imkovic, 
 Rep. Prog. Phys.  {\bf 75}, 106301 (2012).




\bibitem{eji78}
H. Ejiri and J.I. Fujita, 
 Phys. Rep. {\bf 38}, 
85 (1978).


\bibitem{eji00}
H.  Ejiri, 
Phys. Rep. {\bf 338},  265 (2000).
\bibitem{eji19} 
H. Ejiri, J. Suhonen and K. Zuber,
 Phys. Rep. {\bf 797}, 1 (2019).


\bibitem{suh98}
J. Suhonen and O. Civitarese,
  Phys. Rep. {\bf 300}, 123 (1998).
 
 \bibitem{fae98}
 A. Faessler and F. Simkovic,
 J. Phys. G Nucl. Part. Phys.  {\bf 24}, 2139 (1998).
 

\bibitem{suh12}
 J. Suhonen and O. Civitarese,
 J. Phys. G Nucl. Part Phys. {\bf 39}, 085105 (2012).
 
\bibitem{bar13}
J. Barea, J. Korita and F. Iachello, 
 Phys. Rev. C {\bf 87}, 014315 (2013).  
 
 

\bibitem{eng17}
J.  Engel and J. Men\'{e}ndez, 
 Rep. Prog. Phys. {\bf 80}, 046301 (2017).
 


\bibitem{jok18}
L. Jokiniemi, H. Ejiri, D. Frekers, and J. Suhonen, 
Phys. Rev. C {\bf 98}, 024608 (2018).




 
\bibitem{det20}
J. Detwiler, 
 Proc. Neutrino conference 2020  (2020), doi:10.3389/fphy.2018.00160.

\bibitem{eji20}
H. Ejiri, 
 Universe, {\bf 6}, 225 (2020).



\bibitem{eji21a}
H. Ejiri,
 Front. Phys. {\bf 9}, 650421 (2021). 
 
 \bibitem{eji14}
H. Ejiri,  N. Soukouti, and J. Suhonen, 
 Phys. Lett. B  {\bf 729}, 27 (2014).

\bibitem{suh17}
J. Suhonen, 
 Phys. Rev. C {\bf 96}, 055501 (2017). 

\bibitem{suh17a}
J.T. Suhonen, Front. Phys. {\bf 5}, 55 (2017).

\bibitem{eji19a}
H. Ejiri,
Front. Phys {\bf 7}, 30 (2019).

\bibitem{sim13}
F. Simkovic, V. Rodin, A. Faessler and P.
Vogel, Phys. Rev. C. {\bf 87} 045501 (2013).

\bibitem{suh07}
J. Suhonen, From Nucleons to Nucleus: Concepts of Microscopic Nuclear Theory, Springer, Berlin (2007).

 
 




\bibitem{aki97}
H. Akimune, ~et ~al., Phys. Lett. B {\bf 394} 23 (1997), {\bf 665} 424 (2011).
\bibitem{thi12}
J.H. Thies,
~et~al.,   
Phys. Rev. C {\bf 86}, 014304 (2012).


\bibitem{thi12b}
  J. H. Thies et~al.,  
  Phys. Rev. C {\bf 86}, 054323 (2012).


\bibitem{pup11}
  P. Puppe et~al., 
  Phys. Rev. C {\bf 84}, 051305R (2011).


\bibitem{thi12a}
J.H. Thies,
 ~et~al., 
Phys. Rev. C  {\bf 86}, 044309 (2012).

\bibitem{pup12}
P. Puppe, 
  ~et~al., 
Phys. Rev. C {\bf 86}, 044603 (2012).
 
 

\bibitem{eji16a}
H.  Ejiri and D. Frekers, 
  J. Phys. G  Part. Phys. {\bf 43}, 11LT01 (2016).
    
  
\bibitem{aki20}
H. Akimune, et al.,
 J. Phys. G Nucl. Part Phys.  {\bf 47}, 05LT01 (2020).
 

\bibitem{eji19b} 
 H. Ejiri,  
 J. Phys. G Nucl. Part Phys. {\bf 46}, 125202 (2019).
 

 
 \bibitem{eji15} 
H. Ejiri and J. Suhonen,   
  J. Phys. G {\bf 42}, 055201 (2015). 
  
\bibitem{suh07a}
 J. Suhonen, Physics letters B {\bf 607} 87 (2005).



 
 




 

  

\bibitem{hor81}
D.J. Horen et al., Phys. Lett. B  {\bf 99}, 383 (1981).
\bibitem{yos13}
N. Yoshida and F. Iachello,  
 Prog. Theor. Exp. Phys,  {\bf 2013}, 043D01 (2013).


\bibitem{pir15}
P. Pirinen and J. Suhonen, 
  Phys. Rev.  C {\bf 91}, 054309 (2015).

\bibitem{eji17}
H. Ejiri,
J. Phys. G  {\bf 44}, 115201 (2017). 
 
 \bibitem{boh75}
A. Bohr and B. Mottelson,
{\it  Nuclear Structure, II }, W.A. Benjamin, INC., Amsterdam, (1975).

\bibitem{eji89}
H. Ejiri and M. A. deVoigt,
{\it Electron and Gamma -ray Spectroscopy in Nuclear Physics},
Oxford, (1988). 



  
 
 


\bibitem{gys18}
P. Gysbers, et al., 
Nat. Phys. {\bf 15}, 428 (2019). 




\bibitem{coe18}
E. A. Coello Perez, J. Men\'{e}ndez, and A. Schwenk, 
 Phys. Rev. C  {\bf 98}, 045501 (2018). 

\bibitem{sim18}
F. Simkovic, R. Dvornicky, D. Stefanik, and A. Faessler,  
Phys. Rev. C {\bf 97}, 034315 (2018).

\bibitem{cor19}
L. Coraggio, L. De Angelis, T.  Fukui, A. Gargano, N. Itaco, and F. Nowacki, 
 Phys. Rev. C {\bf 100}, 014316 (2019).
 
 \bibitem{men11}
 J. Men\'{e}ndez, D. Gazit and A. Schwenk,  Phys. Rev. Lett. {\bf 107} 062501 (2011).  
 


\bibitem{jok17}
 L. Jokiniemi and J. Suhonen, Phys. Rev. C {\bf 96}, 034308 (2017).
 
 
\bibitem{has18}
I. Hashim, H. Ejiri,  
 T. Shima, K. Takahisa, A. Sato, Y. Kuno,  K. Ninomiya, N. Kawamura, and 
 Y. Miyake, 
  Phys. Rev. C {\bf 97}, 014617 (2018). 
  
\bibitem{jok19}
L. Jokiniemi, J.  Suhonen, H. Ejiri H, and I.  Hashim, 
 Phys. Lett. B {\bf 794}, 143 (2019).  

\bibitem{jok19a}
L. Jokiniemi and J. Suhonen.  
 Phys. Rev. C {\bf 100}, 014619 (2019).    


\bibitem{sim20}
F. Simkovic, R.  Dvornieky, P. Vogel, 
 Phys. Rev.  C {\bf 102}, 034301 (2020).   




 
 




\bibitem{cap15}
F. Cappuzzello, et al., 
  Eur. Phys. J.  {\bf 51}, 145 (2015).

\bibitem{tak17}
K. Takahisa, H. Ejiri, et al., arXiv 1703.08264 (2017).

\bibitem{shi18}
N. Shimizu, J. Menendez, K. Yako, 
Phys. Rev. Lett.  {\bf 120} 142502 (2018).


\bibitem{boh81}
A. Bohr A and B. Mottelson, 
 Phys. Lett. B  {\bf 100}, 10 (1981).

\bibitem{eji82}
H. Ejiri, 
 Phys. Rev. C  {\bf 26}, 2628 (1982).


\bibitem{cat02}
G. Cattapan and L. S. Ferreira, 
Phys. Rep. {\bf 362}, 303 (2002). 

\bibitem{eji21b}
H. Ejiri,  Few-body Syst. {\bf 62}, 60 (2021).



\end{thebibliography}
\end{document}